\documentclass{jfm}
\usepackage{graphicx}
\usepackage{epstopdf, epsfig}

\usepackage{caption}
\usepackage{subcaption}
\usepackage{xcolor}
\usepackage{hhline}
\usepackage{multirow}

\usepackage{amssymb}
\usepackage{amsmath}
\usepackage{amsfonts}       

\usepackage{mathtools}
\usepackage{textgreek}

\usepackage{bm}

\renewcommand{\Re}{\mathrm{Re}}
\newcommand{\Ka}{\mathrm{Ka}}
\newcommand{\Ca}{\mathrm{Ca}}

\shorttitle{On the coupling instability of a gas jet impinging on a liquid film}
\shortauthor{D. Barreiro-Villaverde, A. Gosset, M. Lema and M. A. Mendez}

\title{On the coupling instability of a gas jet impinging on a liquid film}

\author{David Barreiro-Villaverde\aff{1,2}
  \corresp{\email{david.barreiro1@udc.es}},
  Anne Gosset\aff{3}, Marcos Lema\aff{3} 
 \and Miguel A. Mendez\aff{2}}

\affiliation{\aff{1}Universidade da Coruña, CITIC Research, Campus de Elviña, 15071 A Coruña, Spain
\aff{2}von Karman Institute for Fluid Dynamics,
B-1640 Waterloosesteenweg 72, Sint-Genesius-Rode, Belgium
\aff{3}Universidade da Coruña, Campus Industrial de Ferrol, CITENI, 15403 Ferrol, Spain}

\begin{document}

\maketitle

\begin{abstract}
We investigate the dynamics of a gas jet impinging on a thin liquid film. This configuration is relevant to the jet-wiping process and is unstable. In particular, we complement previous works that focused on the wiping of liquids with low Kapitza numbers (highly viscous liquids) by numerically analyzing the wiping of liquids with much higher Kapitza numbers, more relevant to industrial processes. The simulations are carried out by combining Volume of Fluid (VOF) and Large Eddy Simulation (LES), and the dynamics of the gas-liquid interaction is analyzed using extended multiscale Proper Orthogonal Decomposition (emPOD). The resolution and flow details captured by the simulations are unprecedented. The results show that, despite the vastly different wiping conditions, the dynamics of the gas-liquid interaction is remarkably similar. This opens new avenues to the study and the scaling of the jet-wiping process.
\end{abstract}


\section{Introduction}\label{sec:intro}

Gas jets impinging on thin films are used in wiping processes in the coating industry \citep{Buc1997,Gosset2007a}. In these processes, the gas jet acts as an air-knife which controls the thickness of the liquid film deposited on the substrate. The uniformity of the final coating is known to be limited by the instability of the gas-liquid interaction. Long-wave disturbances were first revealed by \cite{Gosset2007}, numerically reproduced by \cite{Myrillas2013}, and \textcolor{black}{extensively} characterized by \cite{Gosset2019} and \cite{Mendez2019}. These works have shown that the primary instability of the wiping consists of large oscillations of the impinging jet, combined with large waves in the liquid film, propagating both downstream and upstream of the wiping region. An extensive analysis of this mechanism was presented by \cite{Barreiro-Villaverde2021}, who combined Volume of Fluid (VOF) and Large Eddy Simulation (LES) with an extension of the multiscale Proper Orthogonal Decomposition (mPOD, \citealt{Mendez2019a}) to study the interaction between the two flows. This investigation revealed that the wiping instability is essentially bidimensional, with waves originating as a result of both displacement of the wiping region and modulation of its strength.

All the aforementioned studies focused on a narrow range of wiping conditions, using low Kapitza number liquids: $\Ka=\sigma \rho_l^{-1} \nu_l^{-4/3} g^{-1/3}$ (with $\rho_l$, $\nu_l$ the liquid's density and kinematic viscosity, $\sigma$ the gas-liquid surface tension and $g$ the gravitational acceleration). These liquids are characterized by a high viscosity and a low surface tension. It implies that the film thickness is relatively large, and thus more ``intrusive'' for the \textcolor{black}{impinging} gas jet. For instance, \cite{Gosset2019} and \cite{Barreiro-Villaverde2021} focused on the wiping of viscous liquids with $\Ka$ in the range $3-5$, typical of continuous painting processes or paper coating. Nevertheless, the much more popular case of wiping in hot dip galvanization is characterized by $\Ka\sim\mathcal{O}(10^4)$ and significantly higher Reynolds numbers in both the liquid film and the gas jet. A detailed investigation of these conditions, however, is still out of the reach of both experimental and numerical fluid dynamics. Studies of these conditions are limited to theoretical models \citealt{Hocking2011,Mendez2020a,Ivanova2022_BLEW3D} or 2D simulations \citealt{Pfeiler2017,Pfeiler2018}. The challenges in high-fidelity simulations of the problem are discussed by \cite{Aniszewski2019}. 

Therefore, an important open question is the extent to which the results obtained for highly viscous fluids are relevant for the much broader spectra of wiping conditions encountered in industry. This short article is an attempt to tackle this question. Specifically, we complement the work in \cite{Barreiro-Villaverde2021} with the investigation of the wiping of a liquid with much higher $\Ka$. Although far from achieving full dynamic similarity with galvanizing lines, as discussed in the following section, the investigated conditions cover a completely different wiping regime than what was previously reported in literature. Therefore, the results give an insight into the dynamics and the scaling of the jet wiping instability, and perhaps enable an educated extrapolation.

\section{Selected Test Cases and Scaling Laws}\label{sec:test_CASE}

We consider planar jet wiping, using a slot nozzle of opening $d_n$ and width $W\gg d_n$, placed at a distance $Z_n$ from a vertical strip moving upwards at a velocity $U_p$. The strip is flat, and the problem is treated as isothermal. The relevant gas and liquid properties are $\rho_g,\nu_g$ and $\rho_l,\nu_l$, respectively, and $\sigma$. The nozzle's stagnation chamber is at a gauge pressure $\Delta P_N$, and the liquid thickness downstream the wiping is $h_f$. The scaling of this configuration is discussed by \citealt{Gosset2019} and \citealt{Mendez2020a}. In addition to $\Ka$, the relevant dimensionless numbers are the Reynolds numbers $\Re_f=U_p h_f/\nu_l$ for the liquid film and $\Re_j=U_j d/\nu_g$ for the gas jet, the dimensionless standoff distance $\hat{Z}_n=Z_n / d_n$, and three numbers. Two of these link the \textcolor{black}{wiping ``strength'' of the impinging jet with the liquid properties. The first is the wiping number ${\Pi}_g=\Delta P_N d_n/(\rho_l g Z_n ^2)$ and relates the maximum pressure gradient (responsible for most of the wiping work) to the liquid density, while the second is the shear number $\mathcal{T}_g=\Delta P_N d_n / (Z(\rho_l g \mu_l U_p)^{1/2})$ and relates the maximum shear stress to the liquid viscosity (see \citealt{Gosset2019}). The third is the dimensionless final thickness $\hat{h}_f=h_f/h_0$, with $h_0=\sqrt{\nu_l U_p/g}$ the maximum thickness that can be withdrawn in the absence of wiping (see \citealt{SPIERS1974389}), and indicates the intensity of the wiping. Finally, the number $h_0/Z_n$ measure the ``intrusiveness'' of the liquid film on the gas jet flow, so as to characterize its geometrical confinement.}


This article analyzes four wiping conditions, for which Table \ref{tab:dless} collects the aforementioned parameters. Two of these (Cases 1 and 2) are taken from \cite{Barreiro-Villaverde2021} and correspond to experiments in \cite{Mendez2019}. They consider the wiping of dipropylene glycol (DG) with a jet of air ($\rho_g=1.2$ kg/m$^3$ and $\nu_g= 1.48 \cdot 10^{-5}$ m$^2$/s) with $Z_n = 18.5$ mm, $d_n=1.3$ mm and $U_p = 0.34$ m/s and discharge velocity $U_j\approx 26 $ m/s (Case 1, with $\Delta P_N=425$ Pa) and $U_j\approx 38 $ m/s (Case 2, with $\Delta P_N=875$ Pa). The other two cases (Cases 3 and 4) were added for this work and consider the wiping of water (W) by an air jet with $Z_n = 10$ mm, $d_n=1$ mm and $U_p = 1$ m/s and discharge velocity $U_j\approx 42 $ m/s (Case 4, with $\Delta P_N=1$ kPa) and $U_j\approx 50 $ m/s (Case 3, with $\Delta P_N=1,5$ kPa). 

It is instructive to compare these wiping conditions with those in a galvanizing line. The table reports the relevant parameters for a moderate wiping of molten zinc by means of an air jet with $Z_n = 15$ mm, $d_n=1.2$ mm and $U_p = 3$ m/s, and discharge velocity $U_j\approx 160 $ m/s ($\Delta P_N\approx 20$ kPa). These would produce a final thickness of about $\approx 25\mu$m. The newly investigated conditions with water are significantly different from previous test cases and closer to galvanizing conditions, as evidenced in Table \ref{tab:dless}.


\begin{table*}
\centering
\caption{Dimensional and dimensionless wiping conditions for cases 1 and 2 with dipropylene glycol (DG), with $\rho_{l}=1023$ kg/m$^3$, $\nu_{l}= 7.33\cdot 10^{-5}$ m$^2$/s, $\sigma_{l} = 0.032$ N/m, Cases 3 and 4 with water (W), with $\rho_{l}=1000$ kg/m$^3$, $\nu_{l}= 1\cdot 10^{-6}$ m$^2$/s, $\sigma_{l} = 0.073$ N/m, and an example of galvanizing conditions with zinc (Z), with $\rho_{l}=6500$ kg/m$^3$, $\nu_{l}= 4.5\cdot 10^{-7}$ m$^2$/s, $\sigma_{l} = 0.78$ N/m. }
\label{tab:dless}
\begin{tabular}{lcccccccccc}
\textbf{Liquid}  & \textbf{Case ID} & $\mathbf{h_f (\mu m)}$ & $\mathbf{h_0 /Z_n} $     & $\mathbf{h_f / h_0} $ & $\mathbf{Re_f}$ & $\mathbf{Ka}$       & $\mathbf{U_j}$ (m/s) & $\mathbf{Re_j}$ & $\mathbf{\Pi_g}$ & $\mathbf{\mathcal{T}_g}$ \\ \hline \hline
\multirow{2}{*}{DG} & 1                                 & 435                    & \multirow{2}{*}{0.085} & 0.27                  & 2.1             & \multirow{2}{*}{4.8}  & 26.6          & 2280            & 0.16   & 0.24            \\
                    & 2                                 & 263                    &                        & 0.16                  & 1.4             &                       & 38.2          & 3274            & 0.33   & 0.41            \\ \hline 
\multirow{2}{*}{W}  & 3                                 & 27.2                   & \multirow{2}{*}{0.032} & 0.085                 & 27.2            & \multirow{2}{*}{3401} & 40.8          & 2690            & 1.02    & 2.23            \\
                    & 4                                 & 21.3                   &                        & 0.067                 & 21.3            &                       & 50.0          & 3297            & 1.53    & 3.02            \\ \hline
\multicolumn{2}{l}{Z}                                   & 25                     & 0.024                  & 0.069                 & 172             & 16440                 & 160           & 12916           & 1.25    & 3.12            \\ \hline \hline
\end{tabular}
\end{table*}

\section{Methodology}\label{sec:methods}

\begin{figure} \center
\centering
\includegraphics[width=0.95\linewidth]{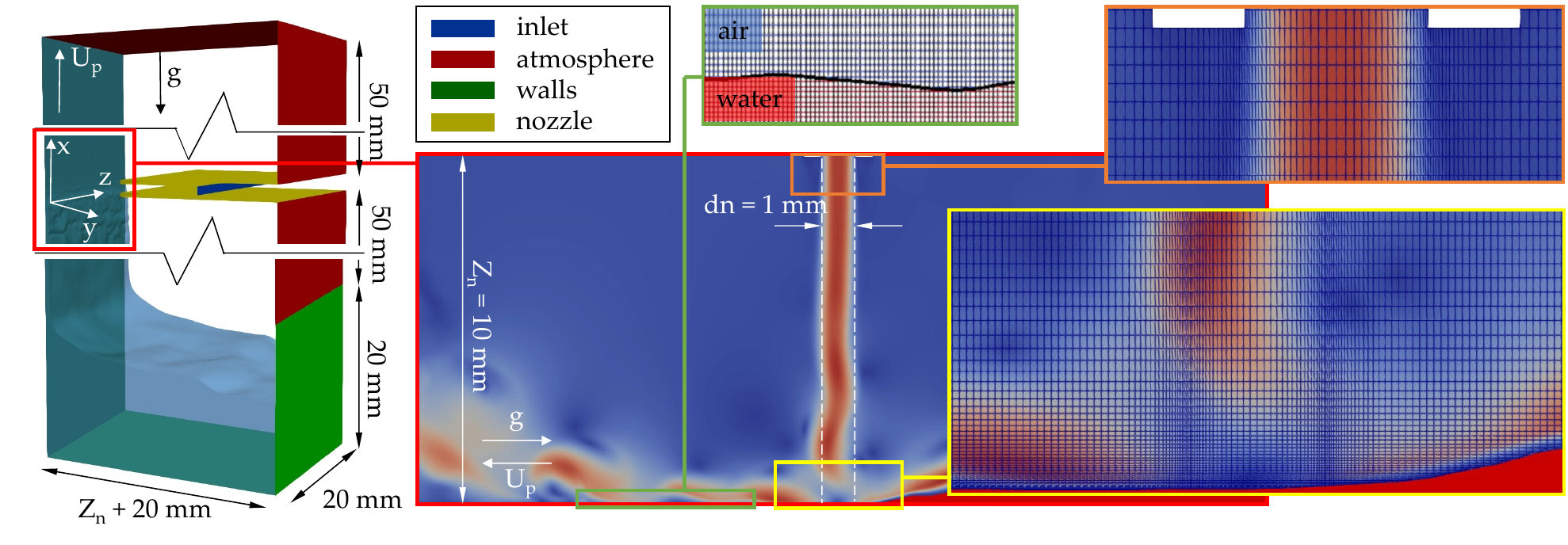}
\caption{Numerical domain, boundary conditions and mesh discretization.}
\label{fig:domain}
\end{figure}

The selected test cases are simulated in OpenFOAM, combining the algebraic VOF formulation of the \textit{interFoam} solver with the Smagorinski model for the turbulence treatment of the gas jet flow. The numerical approach was extensively validated for test cases 1 and 2 in \citet{Barreiro-Villaverde2021}, and is now applied to simulate a much more challenging set-up in cases 3 and 4. 
The domain and snapshots of the mesh (consisting of about 14 million cells) are shown in figure \ref{fig:domain}. The domain includes the coating bath from which the flat substrate is withdrawn, and a portion of the slot nozzle from which the gas jet is released. It spans 50 mm above and below the jet axis in the stream-wise direction ($x$), 20 mm in addition to the standoff distance $Z_n$ in the cross-stream direction ($y$), and 20 mm in the span-wise direction ($z$). The boundary conditions are indicated in figure~\ref{fig:domain}, and the gas jet is established through an inlet condition with prescribed stagnation pressure $\Delta P_N$. The lateral patches (not shown) are set to cyclic.


The discretisation is particularly challenging because of the multiscale nature of the problem. The cell size in the wiping region ($-Z_n<x<Z_n$) is $\Delta_x=50\,\mu$m and $\Delta_y=2\,\mu$m across the film thickness. This provides approximately 10-15 cells within the thickness of the final film, and a minimum of 150 cells per dominant wavelength in the stream-wise direction. The cell size in the span-wise direction is $\Delta_z=250\,\mu$m. On the gas side, the LES index (defined as the ratio between the resolved and the total turbulent kinetic energy) is kept above 80 \% in the wiping region. The simulations are initiated as in \cite{Barreiro-Villaverde2021} and require $\approx$ 100 ms to move past the initial transient, after which 350 ms of physical time are simulated. The time step is $10^{-7}$ s to keep the CFL number below 0.9, and a total of $n_t$=3500 snapshots are exported with a sampling rate of 10 kHz. It requires approximately 1000 hours running in parallel on 512 Intel E5–2680v3 CPUs from the Centro de Supercomputacion de Galicia (CESGA) for one test case with water (case 3).

The results are processed using the extended mPOD as in \cite{Barreiro-Villaverde2021}. In particular, we decompose the film thickness $h(x,y,t)$ into mPOD modes and identify the leading ones as those with the largest and paired amplitudes (denoted as $\sigma_r$), having spatial ($\phi_r(x,y)$) and temporal structures ($\psi_r(t)$) in quadrature. These correspond to travelling waves. The mPOD modes are optimal modes for a prescribed frequency partition (see \citet{Mendez2019a,Mendez2023_chapter}) and have a band-limited frequency content. The temporal structures of the leading waves in the liquid film are then used to identify the most correlated coherent structures in the gas jet. The latter are computed by projecting the velocity fields on the temporal structures $\psi_r(t)$. The reader is referred to \cite{Barreiro-Villaverde2021} for more details on the extended mPOD.

\section{Results}\label{sec:results}

We begin by examining the wave patterns in the liquid film in section \ref{sec:liquid}, and move to the study of the correlated jet structures in section \ref{sec:jet}. Finally, section \ref{sec:wiping} discusses the interaction between the two phases. In all sections, the new results on cases 3 and 4 are analysed along with the ones previously obtained for cases 1 and 2.

\subsection{Wave patterns on the liquid film\label{sec:liquid}}

\begin{figure}
\begin{subfigure}{0.49\textwidth}
  \centering
  \includegraphics[height=4.5cm]{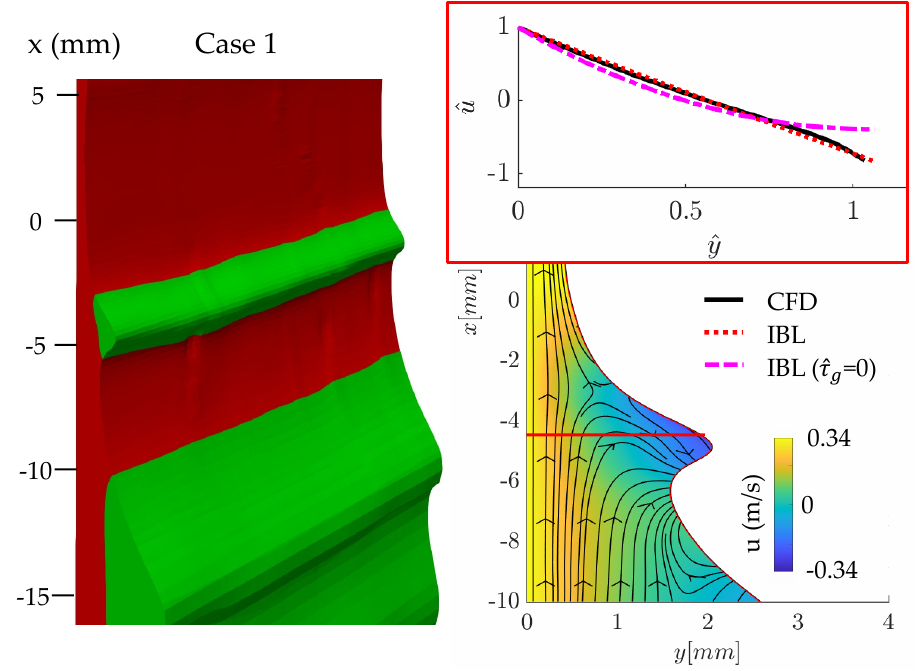}
  \caption{Case 1.}
  \label{fig:runback_1}
\end{subfigure}
\begin{subfigure}{0.49\textwidth}
  \centering
  \includegraphics[height=4.5cm]{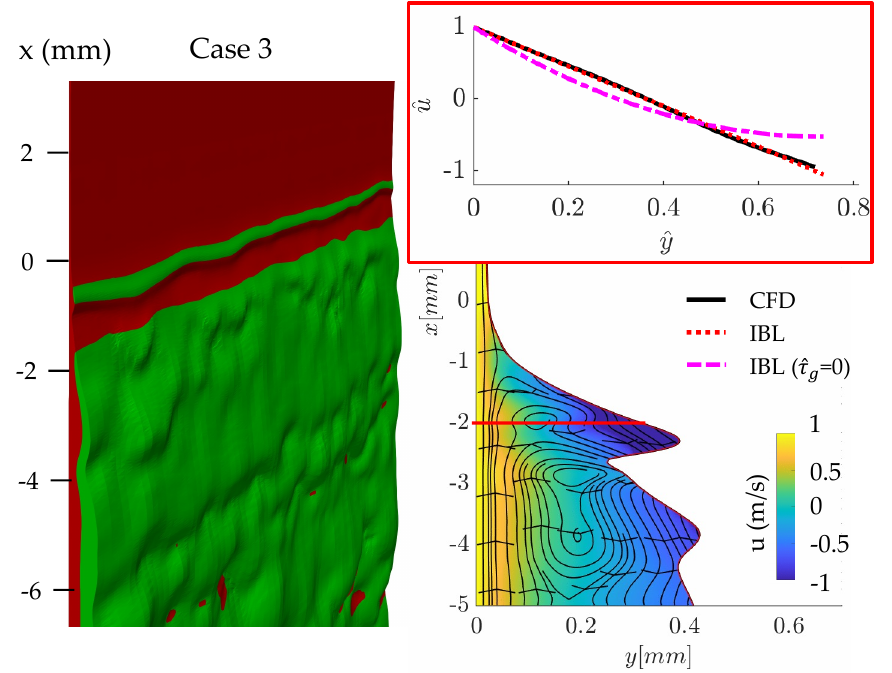}
  \caption{Case 3.}
  \label{fig:runback_3}
\end{subfigure}
\caption{\label{fig:runback} Snapshots of the film dynamics in cases 1 (a) and 3 (b). On the left, a 3D view of the liquid film colored in red where $u>0$ and in green where $u<0$; and on the right, a velocity contour and streamline plot in the midplane. The figure on the top compares the velocity profile at the wave crest with the prediction of the IBL model with and without interface shear stress. A movie of the 3D reconstruction of the liquid film and 2D velocity field taken at the z-midplane is also provided for both cases.} 
\end{figure}

We describe the dynamics of the liquid film upstream the wiping region with the help of figure~\ref{fig:runback}, which compares case 1 (a) and case 3 (b). In each figure, an instantaneous film thickness distribution is shown on the left, with the liquid in red where $u>0$ and in green where $u<0$. On the right, it is complemented with a contour plot of the stream-wise velocity component in the midplane ($u(x,y,z=L_z/2)$), together with three velocity profiles taken at the crest of a run-back wave. These profiles are (1) the one obtained from the CFD computations, (2) the one obtained using the 2D Integral Boundary Layer (IBL) model (equation 3.6 in \citealt{Mendez2020a}) with the flow rate, film thickness and interface shear stress computed from CFD, and (3) the one obtained with 2D IBL model neglecting the gas shear stress. An additional movie of the film thickness distribution with a 2D velocity field taken at the z-midplane is provided for cases 1 and 3. 

In case 1, the liquid film is clearly bidimensional and features large waves originating approximately $2.5$ mm below the jet axis at a frequency of about $20$ Hz, corresponding to $\hat{f}=f \Ca^{-1/3} (U_p \nu_l / g)^{-1/2} \approx 0.11$, with $\Ca=\mu_l U_p/ \sigma$ the capillary number. The reference quantities are taken from the Shkadov-like scaling proposed in \citealt{Gosset2019} for liquid films dragged by moving substrates. These falling waves have a wavelength of the order of $\lambda\approx 12$ mm and evolve over an average film thickness of the order of $h_r\approx 2h_0=$ 3 mm. The Reynolds number in this region, defined as $Re_r=h_r \Delta U/\nu$, with $\Delta U = \max(U_p - u|_{y=h}) \approx 2U_p$ is of the order of $Re_r\approx 18$. The flow is fairly laminar, and the bi-dimensional nature of the falling waves is in line with what happens at the onset of interface stability in falling liquid films. The flow pattern in the film displays a stagnation line at the wavefront, but the assumption of a self-similar parabolic profile seems appropriate. The contribution of the interface shear stress on the velocity profile is much smaller than the viscous and gravitational one, as one might expect from the low shear stress number in this case. 


Case 3 is clearly in a different wiping regime. The Reynolds number in this region is of the order of $Re_r\approx 800$, and the waves rapidly undergo a 3D transition, at approximately $0.5$ mm below the jet axis. 
The velocity field within the liquid is more complex, but several remarkable similarities can be observed. First, the stream-wise velocity component maintains a parabolic shape, even if the influence of the shear stress is more pronounced than in case 1. Second, the frequency of wave formation ($f \approx$ 100-150 $Hz$) in the introduced scaling is $\hat{f} \approx$ 0.15-0.2, a value notably close to the one in case 1. 
Third, and perhaps more importantly for the following discussion, the waves are initially bi-dimensional in both cases. 

\begin{figure*}
\centering
\begin{subfigure}{\textwidth}
  \centering
  \includegraphics[width=\linewidth]{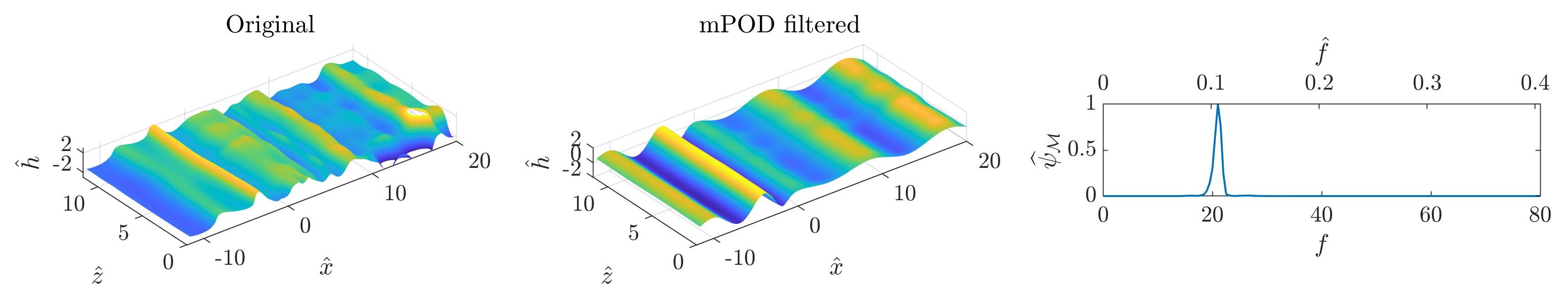}
  \caption{Case 1 (DG):  $\hat{Z}=14.2$,  $\Pi_g =0.16$, $\mathcal{T}_g = 0.24$. }
  \label{fig:mPOD_Z18_P425}
\end{subfigure}\\

\begin{subfigure}{\textwidth}
  \centering
  \includegraphics[width=\linewidth]{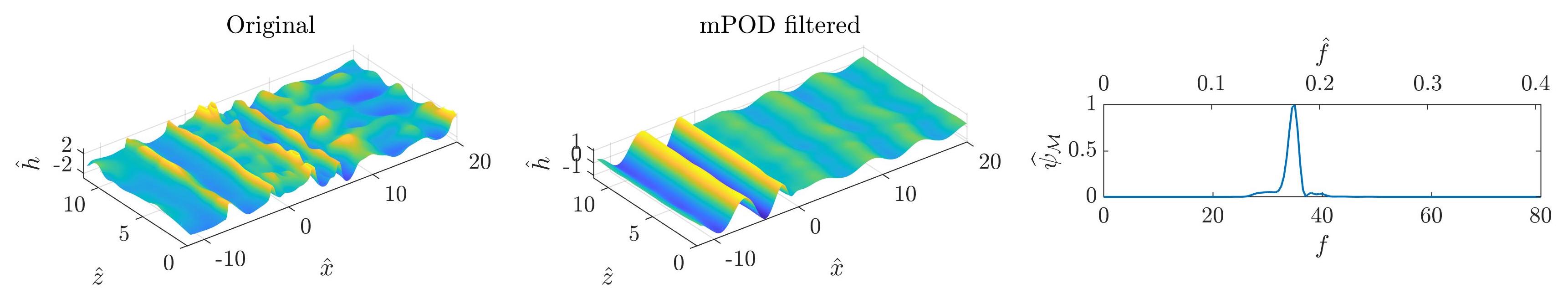}
  \caption{Case 2 (DG): $\hat{Z}=14.2$, $\Pi_g =0.33$, $\mathcal{T}_g = 0.41$.}
  \label{fig:mPOD_Z18_P875}
\end{subfigure}\\

\begin{subfigure}{\textwidth}
  \centering
  \includegraphics[width=\linewidth]{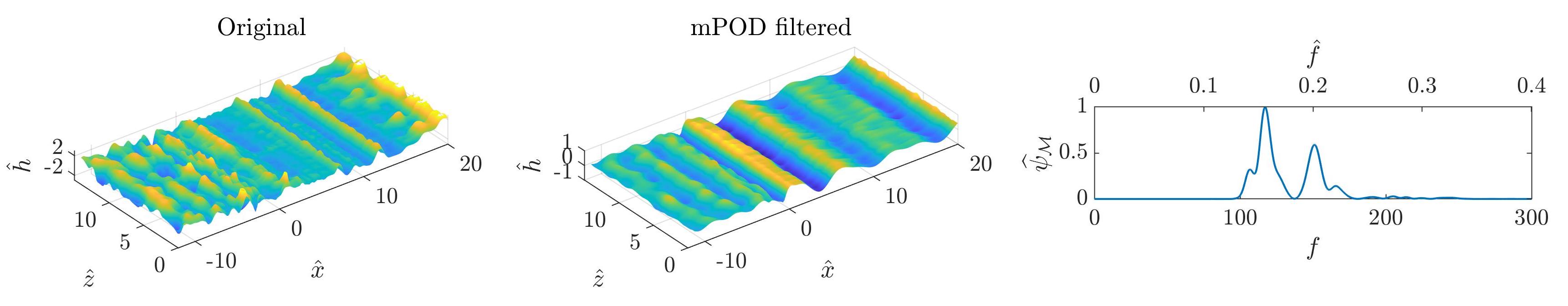}
  \caption{Case 3 (W): $\hat{Z}=10$, $\Pi_g =1.02$, $\mathcal{T}_g = 2.23$.}
  \label{fig:mPOD_Z10_P1000}
\end{subfigure}\\

\begin{subfigure}{\textwidth}
  \centering
  \includegraphics[width=\linewidth]{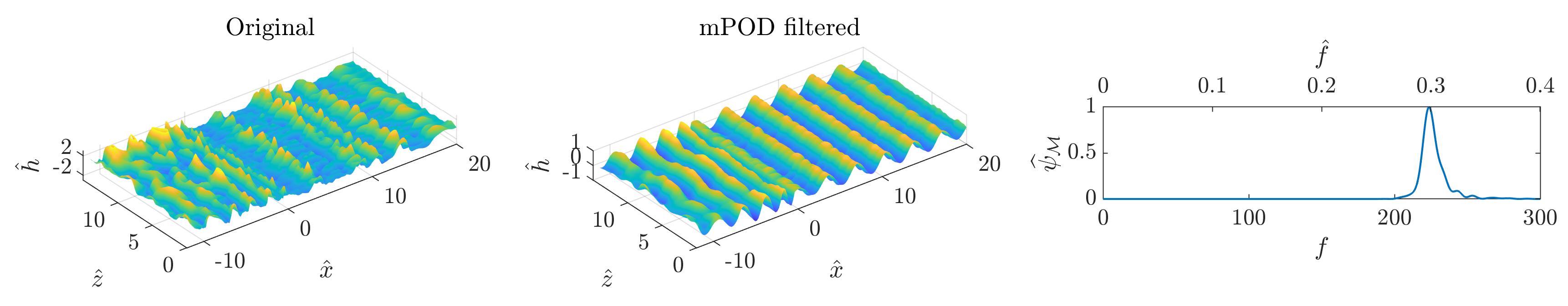}
  \caption{Case 4 (W): $\hat{Z}=10$, $\Pi_g =1.53$, $\mathcal{T}_g = 3.02$.}
  \label{fig:mPOD_Z10_P1500}
\end{subfigure}\\
\caption{Comparison between snapshots of the normalized film thickness $\check{h}(\hat{x},\hat{z})$ (Original) with the leading wave patterns detected via mPOD (mPOD filtered). The spectra of the temporal structure of these modes are shown on the right as a function of the dimensional $f$ and dimensionless $\hat{f}$ wave frequencies.}
\label{fig:mPOD_film}
\end{figure*}

We now move to the liquid film downstream the wiping region. Figure~\ref{fig:mPOD_film} shows snapshots of the liquid film thickness (left), its reconstruction using the leading mPOD modes linked to the dominant travelling wave pattern (middle), and the spectra of the associated temporal structures (right). A snapshot of the four cases is included, and the wiping conditions are recalled in the caption. Both the mPOD analysis and the plots are carried out using the normalized film thickness, defined as $\check{h} (x,z,t) = ({h(x,z,t) - \overline{h}(x,z)})/{\sigma_h (x,z)}$ with $\overline{h} (x,z)$ the average thickness and $\sigma_h (x,z)$ the thickness standard deviation \citealt{Barreiro-Villaverde2021}. This normalization allows the decomposition to equalize the importance of waves downstream and upstream the wiping region despite the largely different thickness. The spectra computed from the temporal structure of the leading modes are shown as a function of the dimensional $f$ and dimensionless $\hat{f}$ wave frequencies.

In all the cases, the leading wave pattern is remarkably two-dimensional in spite of the vastly different wiping conditions, and regardless of the three-dimensional pattern arising upstream the wiping region with water. The dominant wavelength upstream and downstream wiping differs because of the different advection velocities in the two regions, but like in the run-back film, the range of dimensionless frequency is surprisingly similar. In line with the experimental findings on low $\Ka$ liquids in \citet{Gosset2019}, the frequency tends to increase with the wiping strength. On the other hand, the gas jet Strouhal number $St=f Z_n / U_j$ based on the dominant frequency of the waves, is in the range (0.01-0.04), well below the typical values in free-jet instability and hydrodynamic feedback mechanisms \citealt{Mendez2018a}. This suggests a coupling between the gas jet and the liquid, and the goal of the section that follows is to detect the gas jet structures linked to those dominant wave patterns.

\subsection{Gas jet structures correlated with the leading waves in the liquid film} \label{sec:jet}

\begin{figure}
\centering
\begin{subfigure}{0.49\textwidth}
  \centering
  \includegraphics[height=5.4cm]{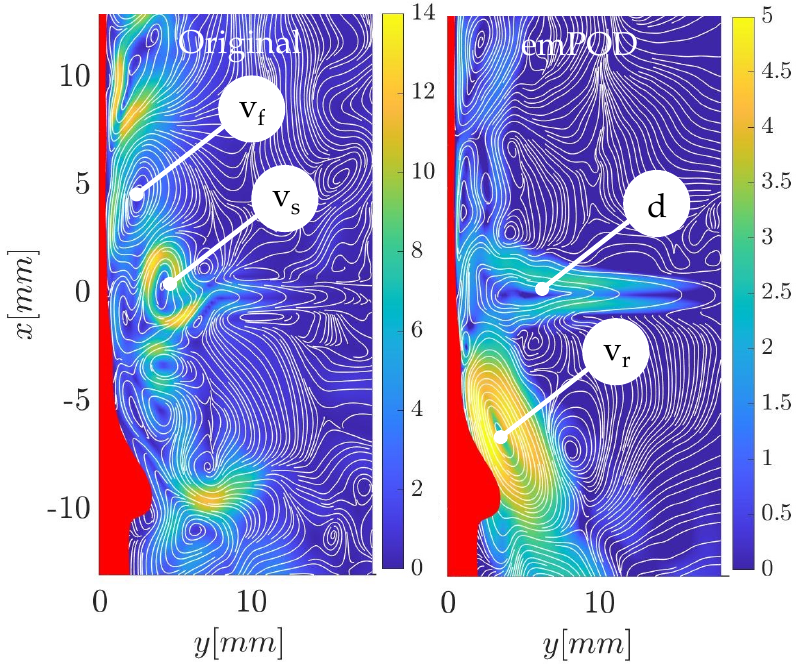}
  \caption{Case 1 (DG)\label{fig:gas_emPOD_case1}}
\end{subfigure}
  \hfill
\begin{subfigure}{0.49\textwidth}
  \centering
  \includegraphics[height=5.4cm]{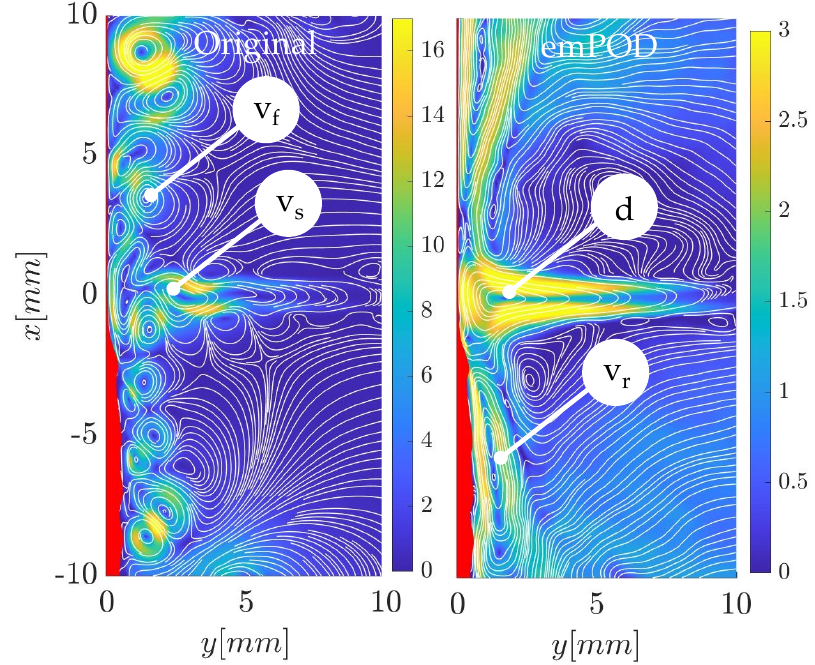}
  \caption{Case 3 (W)\label{fig:gas_emPOD_case3}}
\end{subfigure}
\caption{\label{fig:gas_emPOD} Mean-centred $u^\prime$ and emPOD projected fields for cases 1 (a) and 3 (b). The plots are scaled such that the horizontal axis ranges between $0<y/Z<1$, and the vertical axis between $-10<x/d<10$. \textcolor{black}{A movie for each test case in table~\ref{tab:dless} is also provided.}} 
\end{figure}

We now focus on the spatial structures in the gas jet flow using the extended mPOD (emPOD). This technique allows revealing the flow structures that are most correlated with the leading wave patterns in the liquid film (described in figure \ref{fig:mPOD_film}). Figure~\ref{fig:gas_emPOD} shows two representative snapshots of the gas velocity field $\bm{u}=(u,v)$ and its projection for case 1 (a) and case 3 (b). Cases 2 and 4 lead to similar results and are thus omitted for brevity, but an animation of the film thickness evolution and gas jet flow for each test case is provided in the supplementary material. In each figure, the contour on the left is the mean centred flow field (i.e. $\bm{u}^\prime = \bm{u}(\bm{x}) - \overline{\bm{u}}$) while the one on the right is the projection of the flow field on the leading mPOD modes of the normalized film thickness. 

The original flow fields reveal the complexity of the gas jet dynamics. After impinging the liquid interface, the jet flow splits into two ``side jets'' evolving along the liquid interface. The velocity gradient on the sides of the gas jet core triggers the formation of vortices (label ``$v_s$'') that destabilize the jet near the impingement area and induce small-scale disturbances at the liquid film interface (label ``$v_f$''), which are rapidly damped due to viscuous and capillary damping \citealt{Barreiro-Villaverde2023}. This dynamics, however, is not correlated with the wavy patterns and thus, is not visible in the projected fields on the right. Here, the leading structures consist of a larger vortex induced by the falling liquid waves (label ``$v_r$'') and a rigid oscillation of the impinging jet (label ``$d$''). In the case of low Kapitza liquids, this mechanism was extensively documented in \cite{Myrillas2013, Mendez2019, Barreiro-Villaverde2021}, where a hypothesis on their interaction with the liquid film dynamics was formulated. In particular, it was postulated that the jet oscillation was linked to the vortex-liquid film interaction. The current results with high $\Ka$ liquids show that this interaction is much weaker in these conditions because of the comparatively smaller thickness of the liquid film. Yet, it is shown here for the first time that these structures ``$v_r$'' and ``$d$'' persist, suggesting that the underlying mechanism of the unstable interaction between the gas jet and the liquid film is the same.

\subsection{Dynamics of the gas-liquid instability}\label{sec:wiping}

\begin{figure*}
    \begin{minipage}{.5\linewidth}
      \centering
      Case 1 (DG) \\
      {\includegraphics[width=\linewidth]{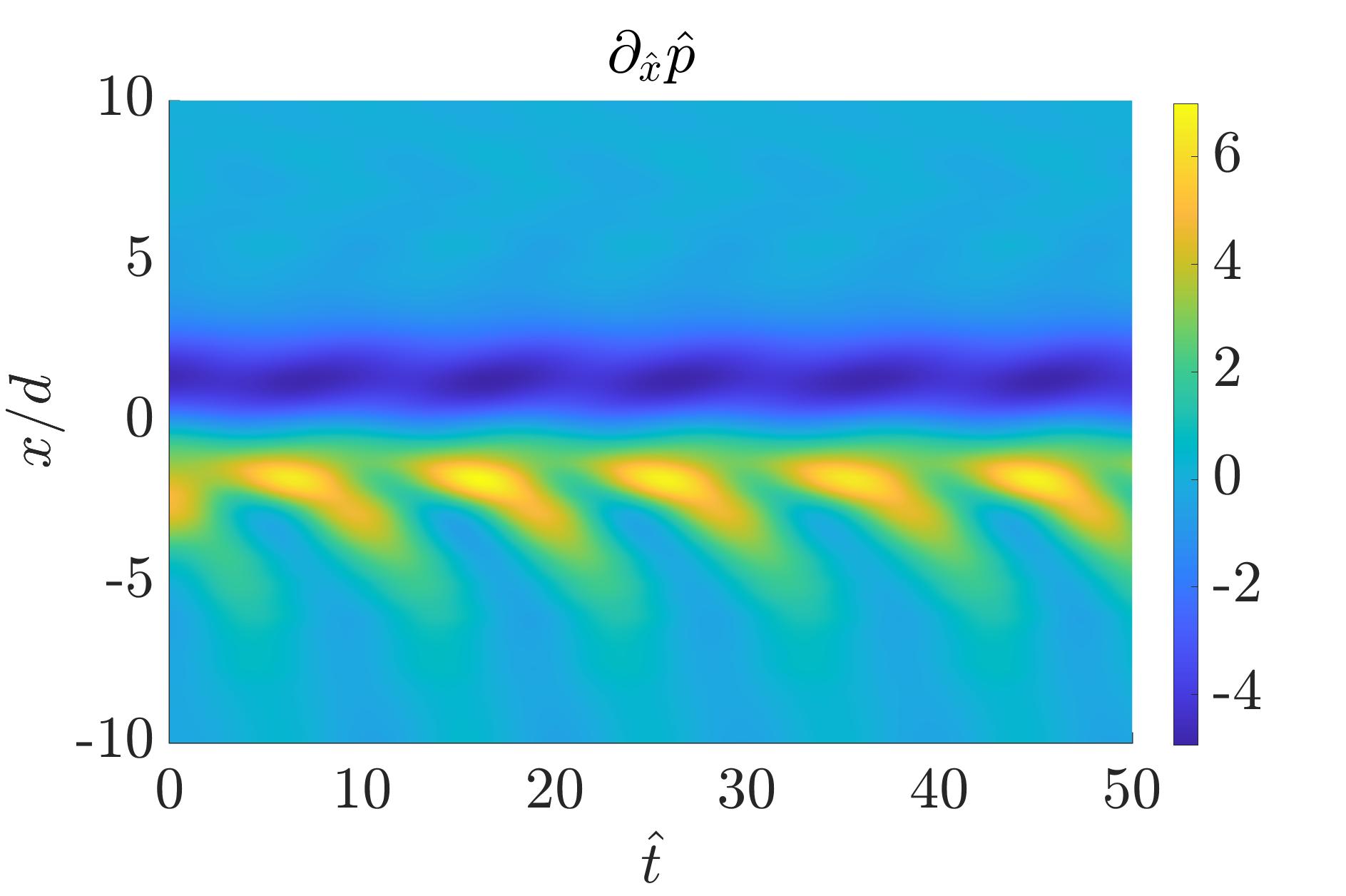}}
      {\includegraphics[width=\linewidth]{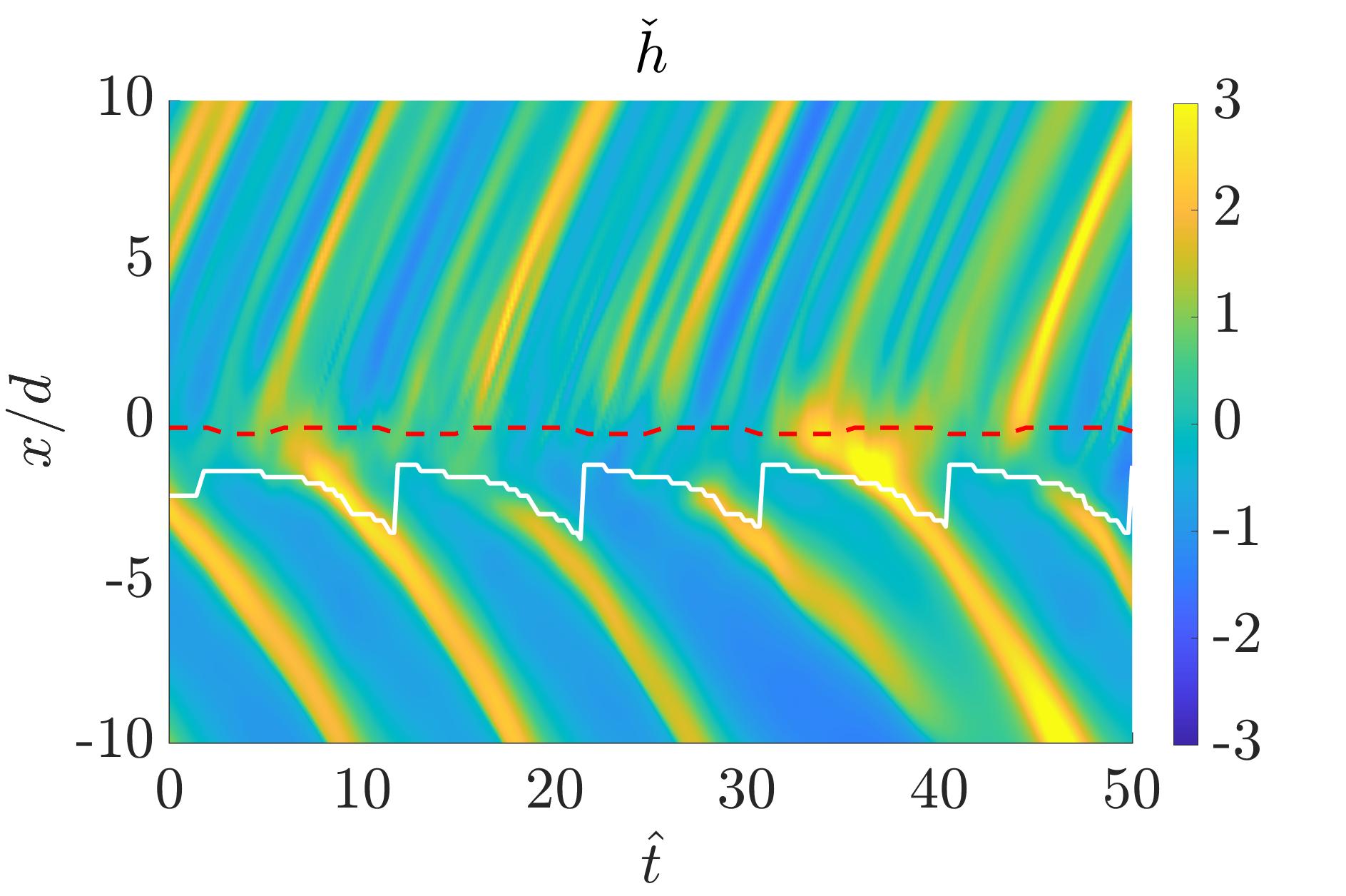}}
    \end{minipage}%
  \hfill
    \begin{minipage}{.5\linewidth}
      \centering
       Case 3 (W) \\
      {\includegraphics[width=\linewidth]{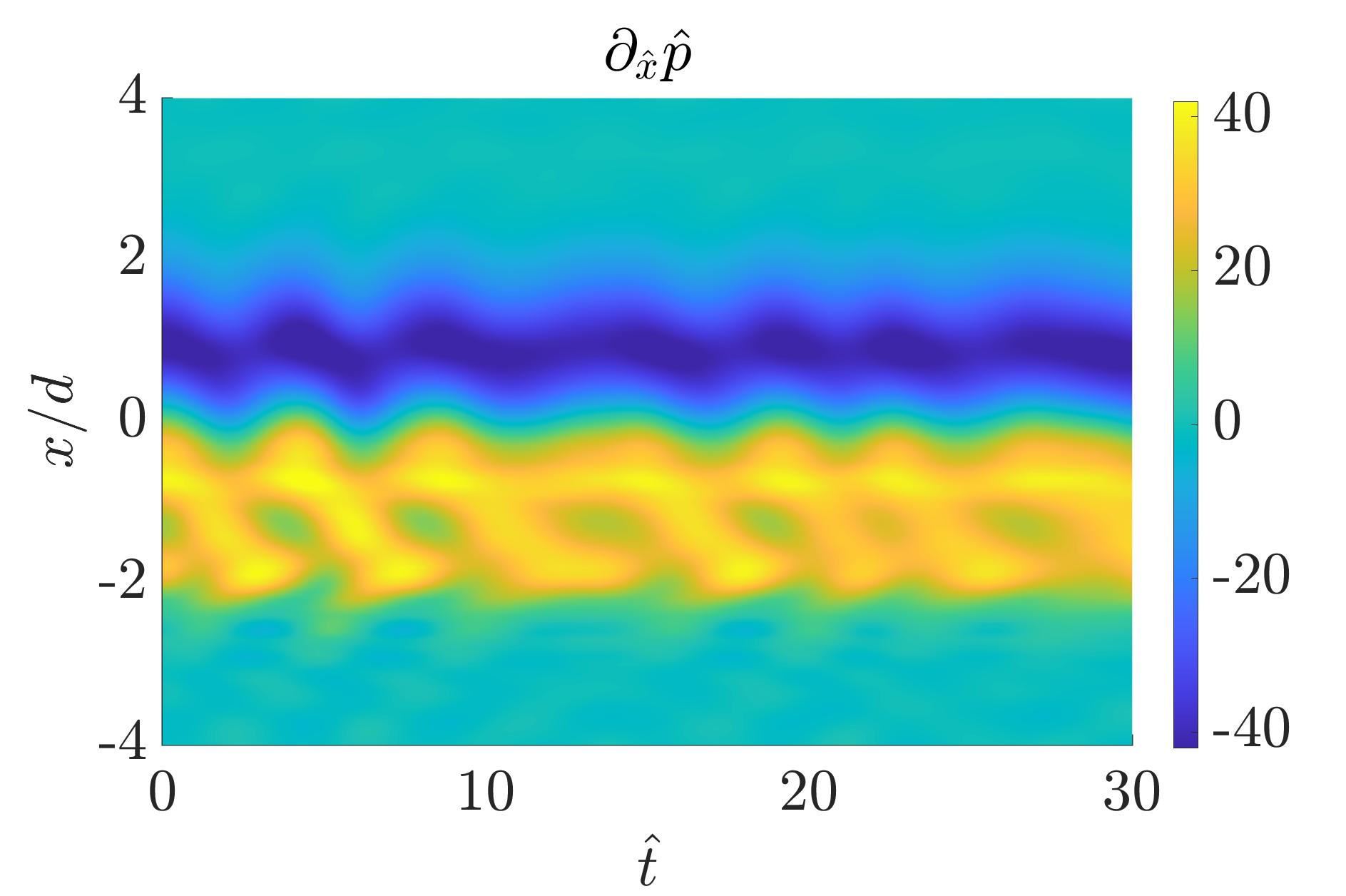}}
      {\includegraphics[width=\linewidth]{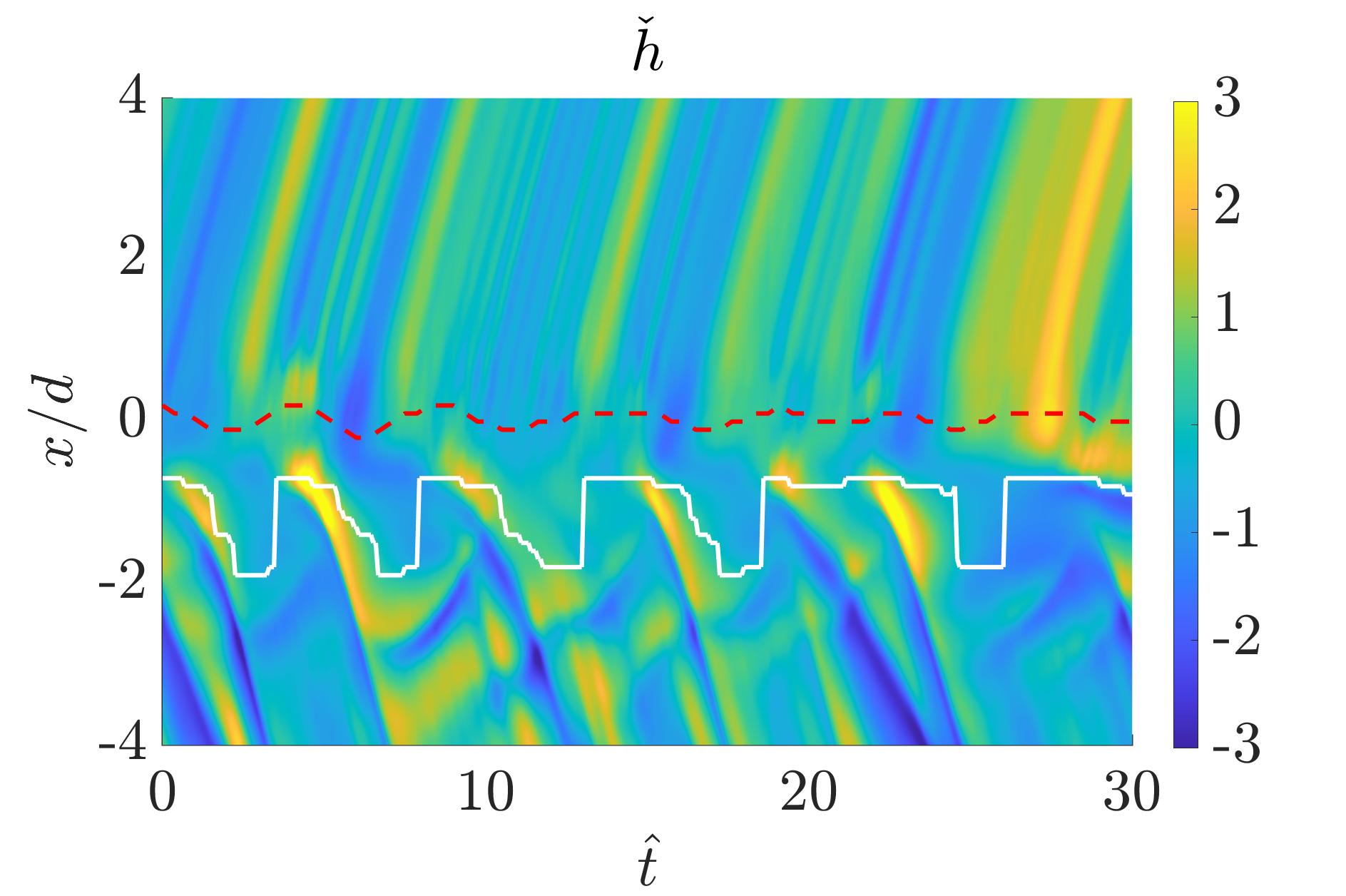}}
    \end{minipage}%
 \caption{Spatio-temporal contours of the dimensionless pressure gradient (top) and normalized film thickness (bottom) for cases 1 (left) and 3 (right). The thickness contours are complemented with the temporal evolution of the impingement point (dashed red line) and the wiping point (continuous white line), i.e. the location of the highest pressure gradient in $x<0$.}
\label{fig:panel_mechanism}
\end{figure*}

Finally, the gas jet-liquid film interaction is analyzed by means of the spatio-temporal contours of the dimensionless pressure gradient $\partial_{\hat x} \hat{p}= \partial_x p/(\rho_l g)$ (top) and normalized film thickness $\check{h}$ (bottom) for cases 1 (left) and 3 (right) in figure~\ref{fig:panel_mechanism}. Both quantities are taken at the mid-plane. The stream-wise coordinate is normalized using the jet opening $d_n$ while the time scale is scaled as in \cite{Mendez2020}, i.e. $\hat{t}= t Ca^{1/3} (U_p \nu_l / g)^{1/2}$. The evolution of the pressure gradient reveals the amplitude of the wiping unsteadiness. The latter is significantly stronger in case 1, resulting in a much larger modulation of the pressure gradient and, thus, of the wiping strength. This difference is clearly linked to the different levels of ``intrusiveness'' of the wiped liquid.

The dynamics of the liquid film is more regular in case 1 than in case 3. Upstream the wiping region, the characteristic lines of the waves are gently curved, but continuous in case 1, while in case 3, waves occasionally coalesce or reverse direction. The space-time contours at the bottom are complemented with a plot of the temporal evolution of (1) the impingement point (dashed red line) and (2) the wiping point (continuous white line), i.e. the location of the highest pressure gradient in $x<0$. In all cases, it appears that the amplitude of the jet oscillation has a minor role compared to the large fluctuations experienced by the wiping point. These fluctuations are characterized by a ``slow'' downward dynamic and a ``fast'' upward dynamic, as if the mechanisms governing these two stages were completely different. The downward shift of the wiping point occurs at the scale of the wave advection time in the falling liquid waves, while the upward shift occurs at the scale of the gas jet advection time. This corroborates the hypothesis that this mechanism is linked to an interaction between the gas jet and the liquid film. The most remarkable result of this analysis is that the same dynamics are observed in both high $\Ka$ and low $\Ka$ cases, even if the wiping conditions and intrusiveness of the film are radically different. \textcolor{black}{Because the dimensionless frequencies of the waves remain within the same range, we may infer that the interaction is mostly driven by the liquid film}. Considering that the dimensionless groups related to the liquid film dynamics are in fair similarity between water and zinc (table \ref{tab:dless}), these results suggest that coupling dynamics observed in water might not differ significantly from the one occurring in galvanization. It is striking that the wavy defects observed on galvanized products have typical wavelengths in the range 10-15 mm, which corresponds to dimensionless frequencies of $\hat{f} = (0.11-0.165)$, totally in line with what is found here for largely different conditions.

\section{Conclusions}\label{sec:conclusions}

\textcolor{black}{We have numerically investigated} the two-phase coupling instability taking place between an impinging gas jet and a liquid film dragged by an upward moving substrate. The two-phase CFD computations cover largely different wiping conditions with a level of detail that is unprecedented in literature. 


It is found that the waves emerge two-dimensional in the impingement region in all the investigated conditions. In some cases, the dominant 2D patterns undergo a 3D transition due to intrinsic instabilities in the run-back flow, or due to the impact of small-scale vortices in the final film. From the gas jet side, the correlation analysis reveals two main structures acting at the time scale of the leading wave pattern: a symmetric oscillation around the jet axis (pattern ``$d$''), and a deflection of the lower side jet triggered by the periodic formation of waves (pattern ``$v_r$''). It is shown that the second mechanism has a stronger impact on the pressure gradient and, thus, on the formation of the waves.



Finally, in spite of the very different flow regimes analyzed in this work, it is remarkable that the dynamics of the gas-liquid interaction is qualitatively similar, and that the wave frequency scales reasonably well using a purely liquid-based scaling. Although the system locks at a certain frequency that depends on both the gas jet and liquid film, these results suggest a dominant role of the liquid film in the coupling.



\bibliographystyle{jfm}

\end{document}